\documentstyle[11pt,newpasp,twoside,epsf]{article}
\def\edcomment#1{\iffalse\marginpar{\raggedright\sl#1\/}\else\relax\fi}
\reversemarginpar

\begin{document}
\title{2MASS Data Mining and the M, L, and T Dwarf Archives}
 \author{J. Davy Kirkpatrick}
\affil{Infrared Processing and Analysis Center, California Institute of Technology, Pasadena, CA 91125, USA}

\begin{abstract}
Studies of brown dwarfs as a distinct galactic population have been 
largely pioneered by dedicated, vast-area
surveys at deep optical and near-infrared wavebands. The Two Micron
All Sky Survey has, because of its full-sky coverage and depth, been the
most successful at providing new targets for further study. The author
briefly reviews some of the ways in which 2MASS data have been used by
brown dwarf researchers. One of the by-products -- namely the
2MASS Team's follow-up spectroscopy of brown dwarf candidates -- 
is potentially valuable as a scientific resource itself and is
also being released to the public.
\end{abstract}

\section{Introduction}

The Two Micron All Sky Survey (2MASS) has, along with the Sloan Digital Sky 
Survey (SDSS) and to a lesser extent the Deep Near-infrared Survey 
(DENIS), provided a treasure trove of information with which to study brown 
dwarfs both in the field and in nearby open clusters. As an illustration,
Table 1 contains a 
breakdown of the number of known nearby L and T dwarfs so far discovered by
each. In addition to providing a discovery tool for field brown dwarfs,
2MASS data in particular have also been used in a variety of
other ways to study substellar objects. The 2MASS
mission is briefly overviewed
in \S2 and some of its contributions to brown dwarf science are highlighted
in \S3.
 
\begin{table}
\begin{center}
\caption{Numbers of L and T Dwarfs Currently Known}
\begin{tabular}{ccc}
\tableline
         & No.\ of     & No.\ of\\
Survey   & L Dwarf     & T Dwarf\\
         & Discoveries & Discoveries\\
\tableline
2MASS    & 155 & 18\\
SDSS     &  72 & 11\\
DENIS    &   9 &  0\\
Others   &   9 &  3\\
\tableline
\tableline
\end{tabular}
\end{center}
\end{table}

Furthermore, Keck Observatory follow-up of 2MASS-selected brown dwarf 
candidates has resulted in a wealth
of high-quality spectroscopic information which the author is releasing to the
general public for the first time. In tandem with the Keck data, the author
is also releasing libraries of other dwarf spectra
that he and his collaborators have amassed over the period 1989-2002. These
libraries are being paired with the 2MASS final release dataset to produce
tables of uniform spectral types and near-infrared photometry for
all dwarf spectral types from K5 through T8. These libraries are further
described in \S4.

\section{Brief Overview of 2MASS}

The Two Micron All Sky Survey is a joint project of the University of 
Massachusetts and the Infrared Processing and Analysis Center at the
California Institute of Technology. As the name implies, the purpose of the
survey was to image the entire celestial sphere at near-infrared wavelengths.
Twin 1.3-meter telescopes were built. One was placed
in North America on Mt. Hopkins in 
southern Arizona, USA, and the second in South America on Cerro Tololo, Chile.
Twin three-channel cameras were also built, one for each telescope, to observe
the sky simultaneously at J, H, and K$_s$ bands using NICMOS 256$\times$256
detectors. The scale was 2{\arcsec}/pixel for a field of view of 
8{\farcm}5$\times$8{\farcm}5. 

The scanning mode for 2MASS was designed to build up 
strips of sky 8{\farcm}5 wide in Right Ascension and 6{\deg} long in
Declination. As the telescope tracked the sky in RA, it would move
at a set rate in Dec. The secondary mirror moved at the same rate in Dec
but in the opposite direction. This motion
would freeze an image of the sky on the
detectors for a 1.3-second integration. After the integration, the secondary
mirror would flip back to its starting position and freeze a new image of the
sky onto the detectors for another
1.3-second integration. Because the telescope continues to move at a constant 
rate in Dec while the secondary is swinging back, this new image is offset
in Dec from the first, in this case by about one-sixth of a frame. 
A sequence of 274 such 
frames would be coadded to give a 6{\deg}-long 2MASS scan where each spot on
the sky is viewed in six consecutive frames for a total integration time of
7.8 seconds. In regions of high galactic latitude, resulting SNR=10 limits of
$J{\approx}16.3$, $H{\approx}15.3$, and $K_s{\approx}14.9$ mag were reached.

By the time these proceedings go to press, 2MASS will have publicly released 
catalogs and images of the entire sky. Details of the release as well
as more information on the 2MASS project itself can be found at the URL 
\verb"http://ipac.caltech.edu/2MASS".

\section{Brown Dwarf Research Enabled by 2MASS}

\subsection{Science by the 2MASS Team}

Initial 2MASS studies of brown dwarfs concentrated on the search and 
characterization of L dwarfs (Kirkpatrick et al. 1999, 2000; Reid et al.
1999) and T dwarfs (Burgasser 2001). Today our team has several other
focuses including the search for M and L dwarfs within $\sim$25 parsecs of
the Sun (see contributions by Cruz and by Wilson), completing a census of
2MASS-detected T dwarfs (see contribution by Burgasser), the search for widely
separated M, L, and T dwarf companions to nearby stars (see contribution by
Gizis), parallax follow-up of confirmed L and T dwarfs from both the northern
and southern hemispheres (see contributions by Harris and by Tinney),
and the analysis of the entire sample for purposes of refining
global properties of brown dwarfs in the solar vicinity (see contribution 
by Liebert).

\subsection{Science by the Community}

Whereas the 2MASS team itself has concentrated on brown dwarfs in the field,
the larger astronomical community has found 2MASS to be a useful tool in the
study of nearby open clusters and star forming regions. Both
Mart{\'{\i}}n et al. (2001) and Brice{\~{n}}o et al. (2002) have used 2MASS
photometry to search for lower mass members of the Taurus star forming region.
Luhman et al. (2000) have used 2MASS to probe the initial mass function in 
Taurus as compared to other star forming regions of differing
star and gas densities. In the Pleiades, Adams et al. (2001) have 
used 2MASS data to determine the extent of the cluster and the rate of
evaporation of its lower mass members. In MBM12, Wolk et al. (2000) 
and Luhman (2001) have used
2MASS to refine the general census of the cluster. In $\sigma$ Orionis,
Oliveira et al. (2002) have used 2MASS data to search for disks.

\subsection{Observational Follow-up of 2MASS Discoveries}

Another avenue of study has been in-depth follow-up of brown dwarfs
and low-mass stars initially selected via 2MASS photometry. Koerner et al.
(1999) acquired conventional 
deep near-infrared imaging around late-M and L dwarfs to
search for cooler companions and to study the frequency and separation ranges
of low-mass binaries. Potter et al. (2002) and Close et al. (2002) have used
adaptive optics to probe even smaller separations. High-resolution
spectroscopic studies for detailed analysis have been obtained by McLean
et al. (2000, 2001) and Nakajima et al. (2001). 2MASS brown dwarfs have also
been scrutinized for photometric variability by Bailer-Jones \& Mundt (1999,
2001), and photometry in other bands -- mainly $V$, $R$, $I$, $z$, $L$, and 
$M$ -- has been obtained by Dahn et al. (2002), Dobbie et al. (2002),
Stephens et al. (2001), and Leggett et al. (2001).

\subsection{Theoretical Follow-up of 2MASS Discoveries}

Other investigators have used 2MASS discoveries to refine their flux
modelling and to derive physical parameters for L and T dwarfs. Burrows et al.
(2000) and Schweitzer et al. (2002) have fit their models to low-resolution 
optical and
near-infrared spectra to derive a temperature scale
for L and T dwarfs. Leggett et al. (2001) have obtained wide wavelength
coverage to compare the observed spectral energy distributions to 
models and to better estimate bolometric luminosities. Geballe et al. (2001)
have obtained new observations of the 2MASS discovery Gliese 570D to determine
a theoretically derived temperature for this very cool T dwarf. High
resolution spectroscopic observations have also been compared to theory by
Schweitzer et al. (2001) to check the validity of the predictions.

\section{The M, L, and T Dwarf Archives}

As they become known, discoveries of L and T dwarfs are served to the 
community through publications and databases such as SIMBAD. 
Although such services are valuable, they fail to present data
in a consistent, homogeneous way since authors do not always present
findings on the same photometric system
or spectral classification system. 2MASS has 
covered the entire sky relatively deeply, so almost all L and T dwarfs -- even
if discovered by other surveys -- will be present in the 2MASS 
images
and catalogs. Furthermore, the 2MASS team has acquired optical spectra of many
of these same objects on a uniform classification system. Hence, it {\it is} 
possible
to present a vast collection of L and T dwarfs with homogeneous photometry and
spectral types.

The author has also obtained many spectra of late-K through late-M dwarfs,
late-type giants, and assorted other stellar spectra in the same wavelength
regime (roughly 6000-10000{\AA}) as that used for optical L and T dwarf
classification. Because of their potential usefulness to the astronomical
community at large, these data will now be served to the public. 
A portion of the archive is shown in Figure 1. The URL for the archives is 
\verb"http://spider.ipac.caltech.edu/staff/davy/
ARCHIVE/." 

\begin{figure}
\plotone{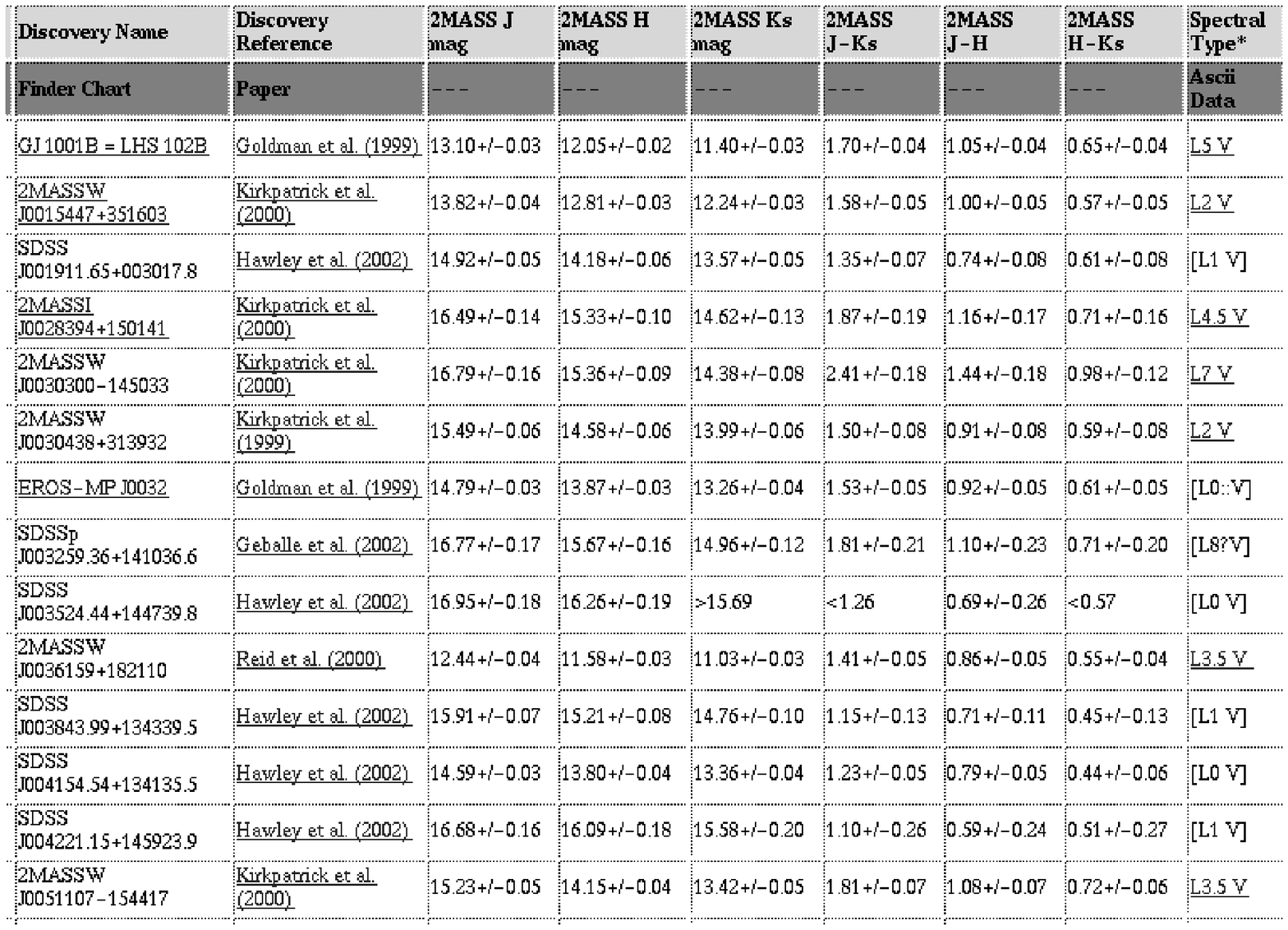}
\caption{A portion of the L dwarf archive as it currently appears.}
\end{figure}

These databases will be disseminated under the aegis of the On-line Archive
Science Information Services (OASIS), a data fusion and visualization applet
accessible through the NASA/IPAC Infrared Science Archive (IRSA). OASIS is
evolving as a portal to data services deployed as part of the National
Virtual Observatory (NVO). Its utility lies in the fact that providers
of small to moderate sized data sets can enable community access to their
data without having to learn NVO standards and protocols.

\subsection{Summary of the Archives}

Table 2 gives a breakdown of data contained in the main archives. The lists
of 245 L and 32 T dwarfs represent a complete collection of all such 
objects known
in the solar vicinity. The lists of 529 K and M dwarfs, 43 late-type 
giants, and 19 M subdwarfs are not meant to be
complete collections in any magnitude- or volume-limited sense
but will nonetheless contain a wealth of data on 
many familiar objects.

\begin{table}
\begin{center}
\caption{Summary of the Archives}
\begin{tabular}{llll}
\tableline
Type of       & No.\ of & No.\ w/ 2MASS & No.\ w/ Optical\\
Object        & Objects & Photometry    & Spectra\\
\tableline
late-K dwarfs &  23 &   23 &  23\\
M dwarfs      & 506 &  506 & 506\\
L dwarfs      & 245 &  242 & 114\\
T dwarfs      &  32 &   29 &  17\\
late-type giants & 43 & 43 & 43\\
M subdwarfs   &  19 &   19 &  19\\
\tableline
\tableline
\end{tabular}
\end{center}
\end{table}

In addition to uniform photometry, uniform spectra, and uniform spectral
types, the archives will also contain positional information, finder charts,
and published
references for each object. Other subsets will also be provided, such
as a list of bright late-K through late-T dwarfs to observe as spectral
standards, a subset of high-SNR spectra with which to compare unclassified
spectra if standards were not obtained at the telescope\footnote{The
author {\bf strongly} recommends the acquisition of spectral standards for
each telescope/instrument combination, rather than relying on comparison
to other researchers' data with different telescopes/instruments or comparing
to uncalibrated spectral ratios.}, and subsets of ultracool dwarfs with 
measured trigonometric parallaxes and absolute magnitudes.

\subsection{Science Already Enabled by the Archives}

Although the archives have existed (in partial states of completeness) for
less than a month, a couple of interesting scientific discoveries have already
been made using the information provided there.

First, in checking the L dwarf archive against areas of sky that 2MASS
imaged more than once, the late-L dwarf 2MASSW J1515008+484742 (discovered by
J.\ Wilson, priv.\ comm.) was found to be moving with a high proper
motion of 1{\farcs}7/yr (R.\ Cutri, priv.\ comm.). For an implied 
spectrophotometric distance of 12 pc, this translates into a
tangential velocity ($v_{tan}$) of $\sim$100 km/s. This is comparable in
size to two of the estimated $v_{tan}$ values in the Gizis et al.\ (2000)
bright L dwarf sample though larger than any of the $v_{tan}$ measures for 
L and T dwarfs in Dahn et al.\ (2002).

Second, during compilation of the archives it was noticed that
the L5 dwarf 2MASSI J0144353-071614 was discovered independently by K.\
Cruz (priv.\ comm.) and by the author and C.\ Tinney (priv.\ comm.). The
spectrum by Cruz -- taken at Cerro Tololo Interamerican Observatory on
24 Jan 2002 -- shows no measurable H$\alpha$ emission (psuedo equivalent 
width pEW $<$ 3 \AA). The
author's Keck spectrum, taken in collaboration with J.\ Liebert
to confirm L dwarf candidates on Tinney's
southern L dwarf parallax program, shows an H$\alpha$ line that is abnormally
strong (pEW$\approx$13 \AA) when compared to other mid-L dwarfs in the 
literature (Gizis et al.\
2000; Kirkpatrick et al.\ 2000). 

\begin{figure}
\plottwo{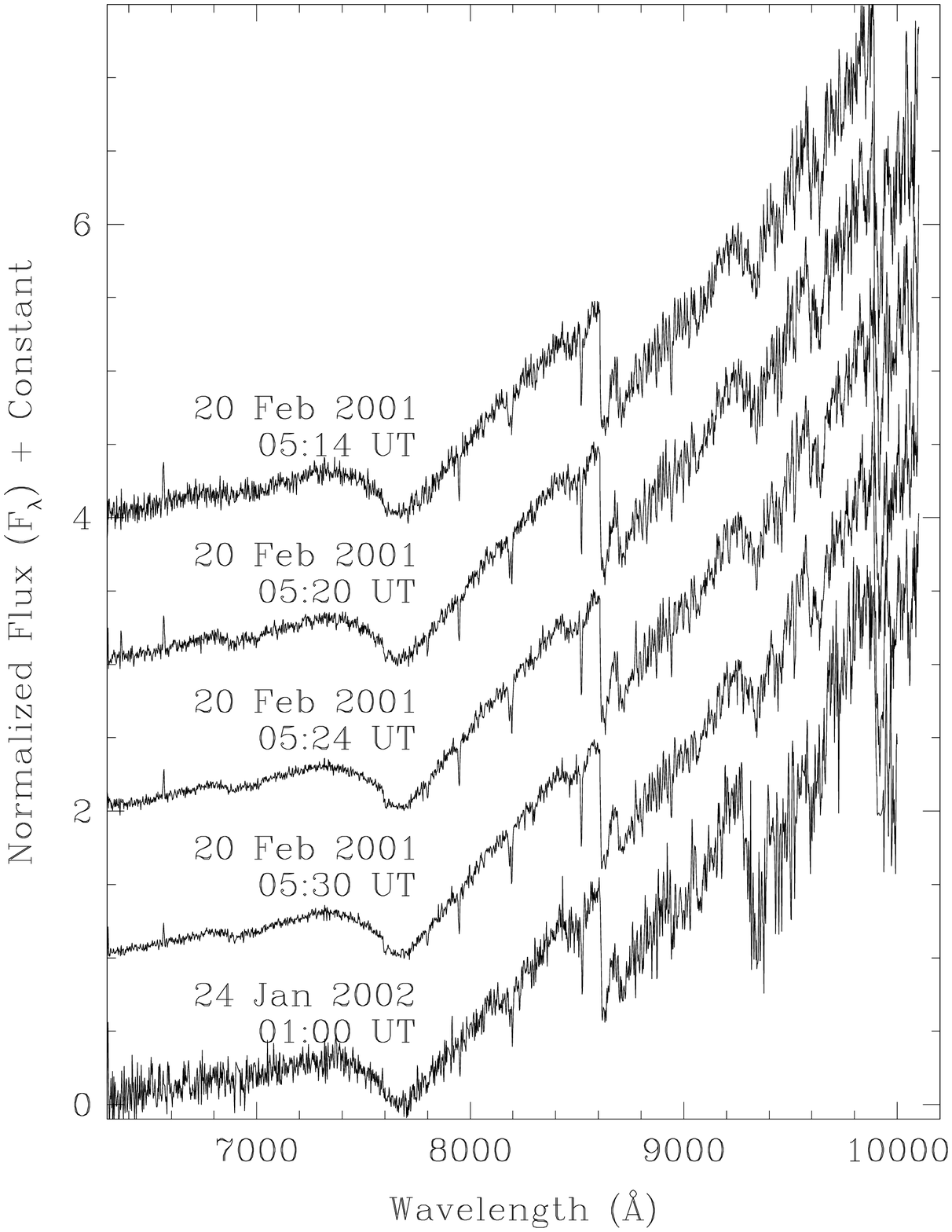}{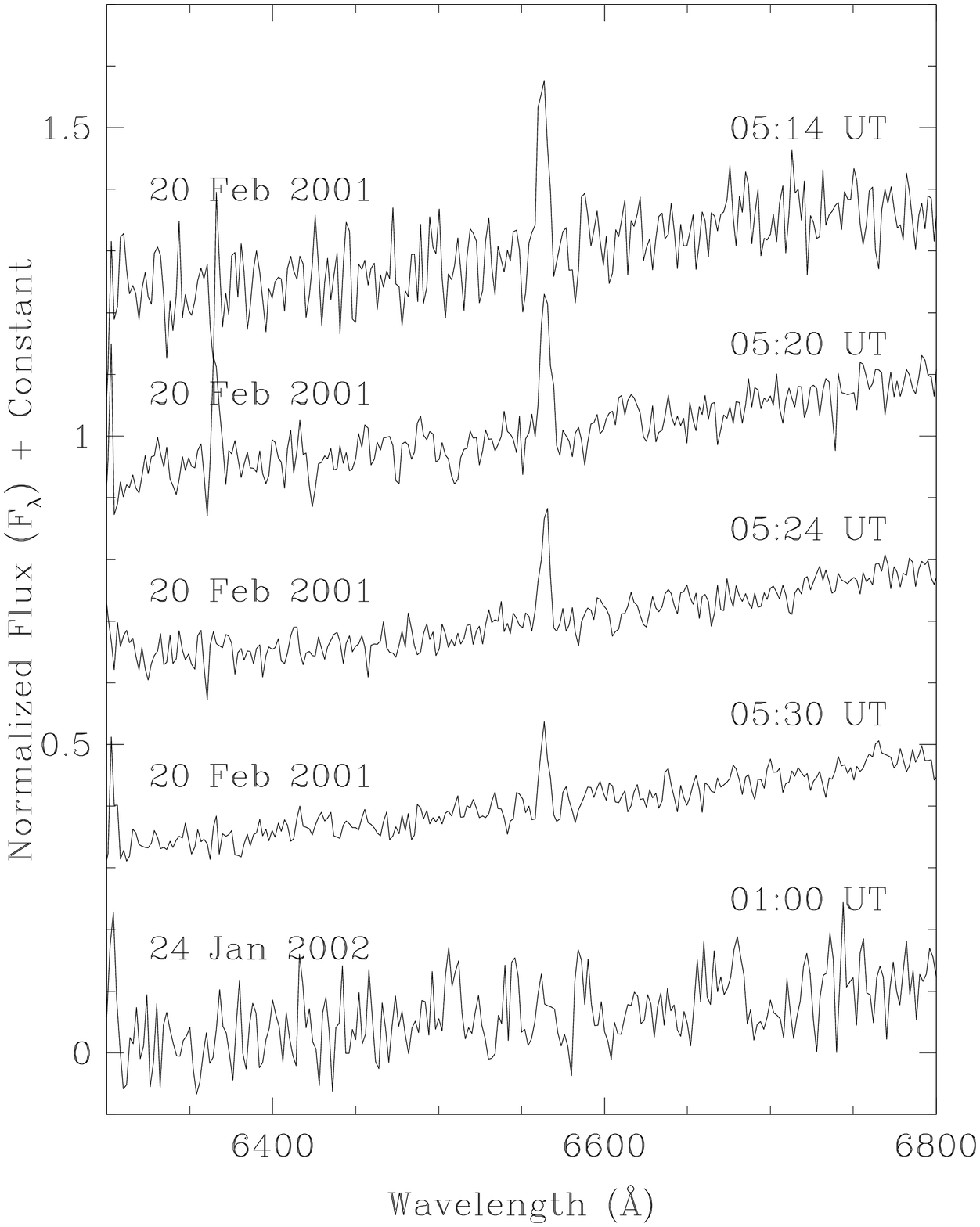}
\caption{In the left figure is shown the sequence of four Keck spectra at
the top and the single CTIO spectrum at the bottom. The four Keck spectra
show an H$\alpha$ line (6563 \AA) that weakens with time, indicating a 
decaying flare event. The right figure shows the same five spectra, this
time zoomed in on the H$\alpha$ line for clarity. Each spectrum has been
normalized to one at 8250 {\AA} and offset by integral (for the left plot)
or half integral (for the right) steps to separate it from neighbor spectra.
No correction for telluric absorption has been applied, which is why the
lower-altitude CTIO spectrum exhibits more 9300-{\AA} water than the Keck
spectra.
}
\end{figure}

This information resulted in a re-analysis
of the original Keck spectrum, which was a coaddition of four separate
exposures. In looking at the individual exposures, it became obvious that
the dwarf was caught declining after a flare event, with the individual
H$\alpha$ lines showing pEW values of 23, 16, 12, and 6 \AA, respectively.
The decline of the flare is shown graphically in Figure 2.
Because Gizis (2002) has cast doubt on the flare event ascribed by Hall
(2002) in the L5 dwarf 2MASSI J1315309-264951, this new observation
becomes the only confirmed flare in an object this cool. As Mullan \&
MacDonald (2001) describe, such a magnetically active object may be proof
that it is, in fact, a star and not a brown dwarf. The lack of a lithium
detection in the Keck spectrum (pEW $<$ 1 \AA) lends some weak support of
this argument. If this is indeed a hydrogen-burning object, it must
be one of the very lowest mass stars.

\vskip 12pt

{\it Acknowledgments:} The author would like to thank John Wilson, Roc
Cutri, Kelle Cruz,
Chris Tinney, and James Liebert for permission to use data
prior to publication.  The author would also like to thank
Patrick Lowrance and Adam Burgasser for
proofreading the manuscript.
This publication makes use of data products from 2MASS, 
which is a joint project of the University of Massachusetts and the 
Infrared Processing and Analysis Center/California Institute of Technology, 
funded by the National Aeronautics and Space Administration and the National 
Science Foundation.
This paper has also made use of the Archive of M, L, and T Dwarfs maintained
at \verb"http://spider.ipac.caltech.edu/staff/davy/
ARCHIVE/". The author wishes to recognize and acknowledge the very 
significant cultural role and reverence that the summit of Mauna Kea has 
always had within the indigenous Hawaiian community.  We are most fortunate 
to have the opportunity to conduct our spectroscopic follow-up
observations from this mountain.

\end{document}